\documentclass[11pt]{article}
\usepackage{acl2016}
\usepackage{times}
\usepackage{url}
\usepackage{latexsym}
\usepackage{graphicx}
\usepackage[multiple]{footmisc}

\aclfinalcopy 

\title{Is your Statement Purposeless? Predicting Computer Science Graduation Admission Acceptance based on Statement Of Purpose}

\author{
\textbf{Diptesh  Kanojia}\textsuperscript{$\dagger$,$\clubsuit$,$\star$}, \textbf{Nikhil Wani}\textsuperscript{$\ddagger$,$\dagger$},  \textbf{Pushpak Bhattacharyya}\textsuperscript{$\dagger$} \\
\textsuperscript{$\dagger$}Center for Indian Language Technology, IIT Bombay, India\\
\textsuperscript{$\clubsuit$}IITB-Monash Research Academy, India\\
\textsuperscript{$\star$}Monash University, Australia\\
\textsuperscript{$\dagger$}\{diptesh, pb\}@cse.iitb.ac.in\\
\textsuperscript{$\ddagger$}nick.nikhilwani@gmail.com\\
}

\date{}

\begin{document}
\maketitle
\begin{abstract}

We present a quantitative, data-driven machine learning approach to mitigate the problem of unpredictability of Computer Science Graduate School Admissions. In this paper, we discuss the possibility of a system which may help prospective applicants evaluate their Statement of Purpose (SOP) based on our system output. We, then, identify feature sets which can be used to train a predictive model. We train a model over fifty manually verified SOPs for which it uses an SVM classifier and achieves the highest accuracy of 92\% with 10-fold cross validation. We also perform experiments to establish that Word Embedding based features and Document Similarity based features outperform other identified feature combinations. We plan to deploy our application as a web service and release it as a FOSS service.

\end{abstract}

\section{Introduction}

Computer Science (CS) graduate admissions process often involves holistic evaluation of  prospective applicant based on multiple subjective and quantitative parameters \cite{ward2006towards}. Amongst these parameters the applicant's Statement of Purpose (SOP) serves as a document to convince its readers' \textit{i.e.} the faculty on the selection committee -  that one has recorded solid achievements which reflect promise for success in graduate study and hence submission of such a good quality SOP becomes of paramount importance.  

Furthermore, Graduate admissions to most Elite universities in the United States of America (USA) only open twice every year - Fall and Spring semesters. 

\textbf{Terminology:} We use the terms essay and SOP interchangeably further during our discussion of the work.

\section{Motivation}

Applicants spend a great deal of time writing SOPs for the admissions process. A well written SOP is a must for an applicant to ensure their admission in any university, and more so for elite universities. Their thoughts and ideas should be organized in their statement. University guidelines\footnote{\url{http://grad.berkeley.edu/admissions/apply/statement-purpose/}}\footnote{\url{http://admission.stanford.edu/apply/freshman/essays.html}}, Alumni blogs\footnote{\url{http://alumnus.caltech.edu/~natalia/studyinus/guide/statement/q&a.htm}}, and Admission consultancy blogs\footnote{\url{http://www.happyschools.com/strengthen-your-graduate-school-application/}} recommend spending ample time on each SOP and tailoring it to perfection. They also recommend stylometry for writing an  essay i.e. word limit, active voice, coherence, and continuity. Various NLP applications like Essay grading \cite{Larkey:1998:AEG:290941.290965}, Text Summarization \cite{gupta2010survey} and Sentiment Analysis \cite{joshi2015harnessing} utilize these features. Hence, we believe that an application that evaluates their statement is crucial. The key question that this paper attempts to answer is:

\medskip
`\emph{
Can information gained from an SOP be used to predict the outcome of a candidate application for graduate school admissions?
}'
\medskip

\section{Related Work}

\newcite{ward2006towards} discuss a qualitative model for Graduate Admissions to Computer Science programs but do not use any Machine Learning or Deep Learning based techniques for estimating a likelihood. According to them, other factors which affect the decision of the committee reviewing the applications include Graduate Record Examinations (GRE) score, Undergraduate Grade Point Average (GPA), Letters of Recommendation (LORs), Financial preparation of a candidate, Alignment with institute needs keeping in mind the diversity goals of the university, and lastly the Undergraduate Major of the candidate. They require the user to rate the application parameters and provide ratings as an input to their system. As an output, they provide an estimate of acceptance based on their model\footnote{\url{http://www.cs.utep.edu/nigel/estimator/}}. 

On the other hand, we employ the existing state-of-the-art techniques, identify features and use some of them to predict the acceptance of a candidate. We acknowledge that we do not model all parameters described above. 

Another similar study \cite{raghunathan2010demystifying} tries to subjectively discuss the admissions process and details the factors which participate in the decision making process of an admission committee. They break the components of a graduate school admissions process and state that SOP is one of the trickiest components of an overall application. They also note that too long an SOP would deter the chances of selection of the candidate. In light of these studies, we focus on creating a model which is able to grade an SOP based on ML techniques.


Text Similarity and related measures \cite{choi2010survey,adomavicius2005toward,gomaa2013survey} have been extensively studied and used for various NLP applications \textit{viz.} Information Retrieval \cite{salton1983extended}, Sense Disambiguation \cite{resnik1999semantic}. To the best of our knowledge, there is no reported study which evaluates SOPs based on the features identified by us, or use ML and DL based techniques of this kind, at the time of submission. Most of the articles list various parameters which are considered by an admissions committee and a Statement of Purpose (SOP) is a common factor among all.

\section{Experiment Design and Setup}

In this section, we provide details about our experiment setup and features used for the classification task.

\subsection{Dataset}

We create our dataset by collecting essays from i) Acquaintances ii) Publicly disclosed SOPs from personal websites, and iii) Admission consultancy blogs. For calculating the similarity measures, we concatenate the essays of the successful applicants, and create a corpus which is used for comparison with both training and testing data. 

We collect a total of 50 manually verified SOPs from Elite Universities (low acceptance rate \textless= 15\%) and rejected essays equally split into two sets. We plan to release the dataset publicly under the CC-BY-SA-4.0\footnote{\url{https://creativecommons.org/licenses/by-sa/4.0/}} license.

\subsection{Methodology}

We use conventional Machine Learning (ML) algorithms \cite{hall2009weka} like Support Vector Machines (SVM) \cite{vapnik2013nature}, Logistic Regression (LR) \cite{walker1967estimation}, and Random Forest Decision Trees (RFDT) \cite{ho1998random} for the task and provide a comparison in Section \ref{sec:results}. 

We use deep learning approaches and deploy a simple Feed Forward Neural Network to classify the SOPs. We split our data in two folds where the first half is used for training, and the second half is then split into tuning and testing datasets. We also use Multilayer Perceptron, another simple Feed Forward Neural Network (NN) and perform a standard 10-fold cross validation on our dataset. We do acknowledge the  modest size of our dataset, but we provide rigorous experimentation including an ablation test to verify that our performance on all classes of our data are unbiased.

\begin{table*}[t]
\small
\centering

\begin{tabular}{|c|c|c|c|c|c|c|c|c|c|}
\hline
Classifier                                                 & $P_{acc}$ & $P_{rej}$ & $\mathbf{P_{avg}}$ & $R_{acc}$ & $R_{rej}$ & $\mathbf{R_{avg}}$ & $F_{acc}$ & $F_{rej}$ & $\mathbf{F_{avg}}$ \\ \hline
RFDT                                                                   & 0.86 & 0.79 & 0.83          & 0.76 & 0.88 & 0.82          & 0.81 & 0.83 & 0.82          \\ \hline
LR                                                                   & 0.69 & 0.83 & 0.76          & 0.88 & 0.60 & 0.74          & 0.77 & 0.70 & 0.74          \\ \hline
SVM                                                                  & 0.89 & 0.96 & \textbf{0.92} & 0.96 & 0.88 & \textbf{0.92} & 0.92 & 0.92 & \textbf{0.92} \\ \hline
\multicolumn{10}{|c|}{\textbf{Neural Network Based}}                                                                                                                    \\ \hline
\begin{tabular}[c]{@{}c@{}}Multilayer Perceptron\\ (Train-Test Split)\end{tabular}          & -    & -    & 0.82          & -    & -    & 0.82           & -    & -    & 0.82          \\ \hline
\begin{tabular}[c]{@{}c@{}}Feed Forward NN (FFNN)\\ (Train-Tune-Test Split)\end{tabular} & -    & -    & 0.36          & -    & -    & 0.60           & -    & -    & 0.45          \\ \hline
\end{tabular}
\caption{Performance of our model on 10-fold cross validation}
\label{tab:results}
\end{table*}

\subsection{Experiment Design}
\label{sec:exdes}

We cluster the set of features in the following groups - \textbf{a) Textual Features} - Feature values based on text contained within the document, \textbf{b) Word Embedding based Features} - Features based on average of vector values provided by pre-trained model on Google News Corpora, \textbf{c) Similarity Score based and Error based features} - Features based on Document Similarity, and other features based on errors in the document. The last set of features have been identified by us, and are our contribution to the work. We, then, use the algorithms mentioned above to calculate precision, recall and F-Score on each feature set.

We also perform an Ablation test to see which feature set combination is performing the best.

\subsection{System Architecture}
\label{sec:sysarch}
\begin{figure}[h]
\centering
\includegraphics[width=1.0\linewidth]{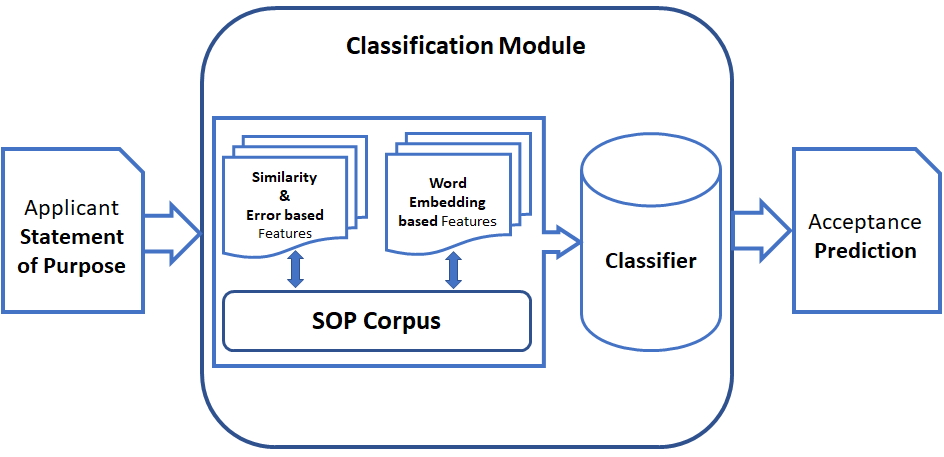}
\caption{System Architecture}
\label{fig:arch}
\end{figure}

Our architecture, shown in figure \ref{fig:arch}, provides the necessary details about the working of our system. The system takes as input the essay of a prospective applicant, calculates feature values for Similarity Score and Error based features along with Word Embedding based features and predicts an \textbf{accept} or \textbf{reject} based on the classification model being used.

\subsection{Features Used}
\label{sec:featu}
We use the following textual features for evaluating the SOPs. These features have been identified via surveying linguistic properties of a text which may affect the organization and quality of an essay.

\subsubsection{Word Embeddings based Features}
\medskip
\begin{enumerate}
\item \textbf{Average Word Vector Scores} - Average of word vectors of each word in the statement calculated using pre-trained Google News word vectors \cite{mikolov2013distributed}.
\end{enumerate}

\subsubsection{Textual Features}
\medskip

\begin{enumerate}
\item \textbf{PoS Ratios} - Ratio of nouns, adjectives, adverbs, and verbs to the entire text, obtained using NLTK\footnote{\url{http://www.nltk.org/}} \cite{loper2002nltk}.
\item \textbf{Discourse Connectors} - It is the number of discourse connectors in the essay computed using a list of discourse connectors\footnote{\url{http://www.cfilt.iitb.ac.in/cognitive-nlp/}}.
\item \textbf{Count of Named Entities} - Number of named entities in the essay. We tried using this as a feature but this drastically lowered the F-scores, and had to be avoided in the final set of reported experiments.
\item \textbf{Readability} - The Flesch Reading Ease Score (FRES) of the text \cite{flesch1948new}.
\item \textbf{Length features} - Number of words in the sentence, number of words in the paragraph, and average word length.
\item \textbf{Coreference Distance} - Sum of token distance between co-referring mentions.
\item \textbf{Degree of Polysemy} - Average number of WordNet \cite{fellbaum2010wordnet} senses per word.

\end{enumerate}

\begin{table}[t]
\small
\centering
\begin{tabular}{c|c|c|c|c|}
\cline{2-5}
 & \multicolumn{4}{c|}{Individual Feature Sets (N-fold)} \\ \hline
\multicolumn{1}{|c|}{Features} & 2-F & 5-F & 10-F & 50\% Split \\ \hline
\multicolumn{1}{|c|}{T  {[}14{]}} & 54 & 46 & 44 & 40 \\ \hline
\multicolumn{1}{|c|}{WE {[}300{]}} & 48 & 78 & 40 & 44 \\ \hline
\multicolumn{1}{|c|}{SE {[}3{]}} & 48 & 56 & 56 & 49 \\ \hline
 & \multicolumn{4}{c|}{Combination of Feature Sets} \\ \hline
\multicolumn{1}{|c|}{T + WE {[}314{]}} & 56 & 62 & 62 & 52 \\ \hline
\multicolumn{1}{|c|}{T + SE {[}17{]}} & 48 & 50 & 38 & 30 \\ \hline
\multicolumn{1}{|c|}{SE + WE {[}303{]}} & \textbf{90} & \textbf{92} & \textbf{92} & \textbf{92} \\ \hline
\multicolumn{1}{|c|}{T + WE + SE {[}318{]}} & 52 & 50 & 53 & 43 \\ \hline
\end{tabular}
\caption{Ablation test on feature sets using Multi-fold Cross Validation } 
\label{tab:resultsAblation}
\end{table}

\subsubsection{Document Similarity Score and Error based Features}
\medskip

\begin{enumerate}
\item \textbf{Cosine Similarity} - Cosine Similarity Score of an SOP with the corpus of accepted essays dataset, where we ensure that the SOP being compared is not a part of the accepted essay corpus. 



\item \textbf{Similarity-based features using GloVe} - The similarity between every pair of content words in adjacent sentences. The similarity is computed as the cosine similarity between their word vectors from the pre-trained GloVe word embeddings \cite{pennington-socher-manning:2014:EMNLP2014}. We calculate the mean and maximum similarity values.

\item \textbf{Spell Check Errors} - We use PyEnchant\footnote{\url{http://pythonhosted.org/pyenchant/}} to embed a spell checker and count the number of errors in each document. The count is then used as another feature for training classifier.

\item \textbf{Out of Vocabulary Words} - We use the pre-trained Google news word embeddings and find out word vectors for every token in the document. The tokens which do not return any vector are either rare words or in all probability out of vocabulary words. We use the count of such tokens as another feature set.
\end{enumerate}

\section{Results}
\label{sec:results}

We perform the experiments detailed in section \ref{sec:exdes} and report our results on 10-fold cross validation. Among the experiments we perform, we achieve the highest F-score of 92\% using the SVM classifier with an RBF Kernel. The results are shown in table \ref{tab:results} and discussed in Section \ref{sec:disc}.

Table \ref{tab:results} clearly indicates that SVM outperforms Random Forest Decision Trees (RFDT) with a margin of 9\%, Logistic Regression (LR) with a margin of 18\%, Neural Network based Multilayer Perceptron with a margin of 10\%, and another Feed Forward Neural Network (FFNN) with a margin of 47\%. We further discuss the impact and justifications of these results in Section \ref{sec:disc}.

We also perform a multi-fold ablation test, using SVM Classifier, on the feature sets identified in section \ref{sec:exdes}. The results for the ablation test are shown in Table \ref{tab:resultsAblation}. The table clearly identifies that Similarity Scores and Error based features along with Word Embedding based features give us the best results.

\section{Discussion}
\label{sec:disc}

In order to identify the features that contribute to the modeled non-linearity of SVM and our best reported accuracy of  92\%, we conduct a comprehensive ablation test. Feature sets mentioned in Section \ref{sec:exdes} were considered. A total of 317 features were ablated based on their sets via multi-fold stratified cross validation experiments and additionally in an experiment with 50\% split of the dataset as shown in the Table \ref{tab:resultsAblation}. 

It was found that the 14 identified Textual (T) features  do not contribute significantly to our model. We extrapolate that these features may have worked better in another context such as Sentiment Analysis \cite{mishra2017leveraging}, or Essay Grading \cite{valenti2003overview}, but not for the task of SOP Classification. Our task primarily aims at labeling an SOP with an accept or reject, however, we observe that Textual features do not differentiate well between coherent and incoherent essays. We also observe that Word Embedding (WE) features of 300 dimensions contribute significantly towards the accuracy of our final model. While they do not contribute notably when used to perform classification independently, combining them with Similarity Score and Error Based (SE) feature set form our best reported model i.e. SE + WE.

\section{Conclusion and Future Work}

In this paper we demonstrate the applicability of a data driven approach to mitigate the unpredictability of Computer Science graduate admissions process. We build a corpus of fifty manually verified SOPs from Accepted applicants to Elite Universities (low acceptance rate \textless= 15\%) rejected SOPs. We show that a combination of Cosine Similarity, Error based features and Word Embedding based features outperform any of the textual features based combinations, for this task. Based on the ablation tests conducted, we model an SVM classifier that predicts with significantly high accuracy. 

In future, we plan to integrate Parts-of-speech (POS) based similarity measures and Recurrent Neural Networks (RNN) \cite{cho2014learning} which have been shown to work well with textual data. Integration of other traditional metrics of a candidate’s application performance measure such as GRE, Test of English as a Foreign Language (TOEFL) / International English Language Testing System (IELTS) score and GPA will further robustly extend this model. We also plan to translate this novel research to an open source web application which would allow prospective applicants to evaluate their SOPs with our system.

\bibliography{acl2016}
\bibliographystyle{acl2016}

\end{document}